# Analysis of Decentralization in Governance and Financial Efficiency of Companies: Studying the Relationship in the Field of Decentralized Finance

Kolmykov K.A.[1]

## Abstract

Currently, the advantages of decentralization through blockchain technology in the financial sector are actively discussed. In this article, we investigate the decentralization in the governance of Decentralized Autonomous Organizations (DAO) using the Gini coefficient as an indicator of inequality among the token owners. This metric is analyzed in the context of Return on Investment (ROI) for companies in the decentralized finance (DeFi) sector. Our goal is to understand whether the level of "real" decentralization in blockchain-based governance affects financial efficiency, and to explore the benefits and possible limitations of such an approach. This analysis allows for a deeper understanding of the significance and impact of decentralization on the functioning and productivity of organizations in the DeFi sector, and to determine the extent to which this impact is positively or negatively reflected in their success and profitability. Additionally, the results of this analysis will provide a fuller understanding of the dynamics and potential of blockchain for organization governance.

---

[1] Student of the Faculty of Economics of Lomonosov Moscow State University.

## Introduction

The modern world is witnessing a rapid development and implementation of digital technologies. In recent years, the blockchain technology, which is finding increasing applications in various sectors of the economy, has piqued significant interest. Our focus is on the application of blockchain in the realm of finance, specifically decentralized finance (DeFi). DeFi is gaining more popularity and offers new opportunities for creating financial products and services. This sector, in turn, has given rise to Decentralized Autonomous Organizations (DAO).

The uniqueness of DAOs lies in the utilization of blockchain technology to create an organizational structure that doesn't require centralized governance. This enables the creation of entirely new governance models and ways of interaction between stakeholders. However, how effective are these models? Does "real" decentralization influence the financial efficiency of companies in the DeFi sector?

In this article, we present a study aimed at understanding the relationship between decentralization in DAO governance and the financial efficiency of these companies in the DeFi realm. The analysis is based on data from 32 companies operating in the DeFi sector. Key metrics used in the study include the Gini coefficient (a measure of token distribution inequality) and Return on Investment (ROI).

Our goal is to get a clearer picture of the impact of decentralization on financial efficiency, and to understand whether it can be a factor in enhancing investment attractiveness and ensuring sustainable growth of companies in the DeFi sector. Hence, this research will pave the way for further academic endeavors in this direction and stimulate the community of researchers and practitioners to continue exploring this promising area.

## Literature Review

### 1.1. Blockchain Technology and its Application in Finance: From the Traditional Financial Sector to DeFi

In today's global economy, which is grounded in information and knowledge, digital ecosystems play a pivotal role. These interconnected networks of businesses, individuals, and organizations collaborate and exchange value through digital technologies. Blockchain technology has become the foundation for creating decentralized digital ecosystems, ensuring secure, transparent, and efficient value exchange without the intermediaries traditionally associated with the financial sector.

The financial sector, comprising all corporations and quasi-corporations involved in delivering financial services, has shown particular interest in blockchain. Increasingly, attention within this sector is given to financial technologies, or fintech. Fintech encompasses the use of innovative technologies to enhance and optimize traditional financial services.

In the publication titled "Financial Technologies in the Banking Industry: Challenges and Opportunities", a fintech industry map is presented, divided into eight broad categories such as: payments, insurance, planning, lending/crowdfunding, blockchain, trading & investment, data & analytics, security (Ajlouni, Al-Hakim, 2018). Among these, blockchain stands out as a significant innovation due to its origin story and its core operational principles.

While blockchain was initially developed to support financial transactions, its application today extends far beyond the financial sector. Increasingly, various industries are harnessing this technology to improve the efficiency and security of their processes.

According to the work titled "Blockchains and the Economic Institutions of Capitalism", blockchain possesses significant potential for application in institutional economics, offering a novel way of coordinating economic activity. Indeed, blockchain embodies many key institutional aspects of market capitalism: property rights, exchange mechanisms, money, law (code), and finance (initial coin offerings) (Davidson et al. 2017).

The cryptocurrency market, initiated by blockchain, has rapidly gained popularity over recent years. Initially, cryptocurrency trading took place via centralized exchanges, like Binance (categorized as CeFi). However, with the evolution of blockchain technology and smart contracts, there arose a demand for more sophisticated financial applications offering greater flexibility. This led to the emergence of decentralized finance (DeFi) - an ecosystem of financial applications built on blockchain networks and utilizing smart contracts. DeFi continues to push the boundaries of financial services, offering new and innovative solutions for consumers and organizations.

## 1.2. Decentralized Finance (DeFi)

The 2022 annual report from the analytical agency CoinGecko categorizes DeFi into the following sectors: DEXs (Decentralized Exchanges), Oracles (information conduits),

Lending, Derivatives, Liquid Staking, Yield Aggregators, Insurance, Asset Management, and Fixed Interest (CoinGecko, 2022).

These sectors can be broadly classified into two main groups based on their business models:

1. Novel to the financial sector, wholly crafted due to the capabilities of blockchain and smart contracts.
2. Dependent on or rooted in traditional finance, which have been adapted for incorporation into DeFi.

Blockchains that support smart contracts and smart contract platforms provide the foundation upon which applications are built, but they aren't separate sectors within DeFi. The list of DeFi sectors represents specific groups of applications or services that exist and function thanks to this infrastructure. Hence, within our analysis of DeFi sectors according to CoinGecko, blockchains and smart contract platforms are considered as the basic technological support, not as distinct sectors.

In this chapter, we conduct a comprehensive analysis of each of these sectors, drawing upon data from existing DeFi organizations, and present tables (Table 1,2) showcasing their characteristics.

Table 1 features business models of DeFi organizations classified as having no prior analogs in the financial sector. Table 2 displays "Business models based on traditional finance," and the corresponding financial corporations, based on the categories of financial corporations (FCs) mentioned in the "Monetary and Financial Statistics Manual and Compilation Guide" by the IMF.

**Table 1.** Group 1: Business Models Wholly New for the Financial Sector

| Sector | Description | Example Organizations |
|---|---|---|
| Decentralized Exchanges (DEXs) | Decentralized exchanges enable peer-to-peer (P2P) trading by leveraging one of cryptocurrency's fundamental capabilities: facilitating financial transactions without intermediaries. | Uniswap, SushiSwap, Tinyman |
| Yield Aggregators | Platforms that automate the process of finding strategies and directly investing in the most efficient DeFi protocols to maximize returns. | Yearn.Finance, DeFi Yield |

| | | | |
|---|---|---|---|
| Liquid Staking | | Solutions that allow users to provide liquidity and earn additional income, while simultaneously using these tokens within the DeFi ecosystem. | Lido, Rocket Pool |

**Table 2.** Group 2: Business Models Based on Traditional Finance

| Sector | Associated Financial Corporations (FCs) | Explanation | Example Organizations |
|---|---|---|---|
| Oracles | Auxiliary Financial Corporations [3.166] | Oracles provide data for smart contracts, acting as financial auxiliary corporations that facilitate the exchange of information and services in traditional finance. | Chainlink, Band Protocol |
| Lending | Other Deposit Corporations (ODCs) [3.123], Corporations accepting deposits except the central bank [3.124] | Lending platforms offer borrowing and lending services, akin to traditional deposit-taking institutions, such as banks. | Aave, Compound, MakerDAO |
| Derivatives | Investment Funds, not linked with MMF [3.149], Auxiliary Financial Corporations [3.166] | Derivative platforms allow trading of financial instruments, similar to investment funds not linked with MMF and auxiliary financial corporations facilitating the trading of derivative financial instruments. | dYdX, Synthetix, GMX |
| Insurance | Insurance Corporations [3.190] | Insurance platforms offer risk coverage, similar to traditional insurance corporations. | Nexus Mutual, Cover Protocol |
| Asset Management | Investment Funds, not linked with MMF [3.149], Auxiliary Financial Corporations [3.166] | Asset management platforms simplify the management of a crypto portfolio, similar to traditional investment funds not linked with MMF and auxiliary corporations. | SwissBorg, Enzyme |
| Fixed Interest | Other Deposit Corporations (ODCs) [3.123], Corporations accepting deposits except the central bank [3.124], Auxiliary Financial Corporations [3.166] | Fixed-interest platforms offer fixed deposit interest rates, reminiscent of traditional deposit-taking institutions and auxiliary financial companies providing debt financial instruments. | BondAppetit, Anchor |

The development of DeFi has also led to the creation of new digital assets specifically designed for use in decentralized applications. These assets, known as "DeFi tokens," are becoming increasingly popular among investors and traders, aiding in the stimulation of both the DeFi and cryptocurrency markets.

Overall, the emergence of DeFi represents a significant evolution in the cryptocurrency market, blockchain technology, fintech, and the financial sector. It expands the potential applications of blockchain beyond simple currency transfers to encompass a wide range of financial services. Consequently, traditional financial institutions are compelled to become more technologically advanced to maintain their competitiveness, not just against other financial entities but also against emerging fintech companies (Aleshina, 2022). Collectively, these factors could potentially democratize access to financial services and provide greater financial freedom to those who need it.

## 1.3. Decentralized Autonomous Organizations (DAO)

The concept of a Decentralized Autonomous Organization (DAO) represents an expansion of blockchain technology's horizons, which gained traction with the growth of the cryptocurrency market. Specifically, the implementation of smart contracts in Ethereum paved the way for its development and popularization.

In the financial domain, the emergence of DAOs has ushered in a new chapter in the evolution of governance structures within digital blockchain ecosystems. DAOs, for the most part, pertain to such innovative decentralized structural models as "federative" and "communal". Their business models often take the form of decentralized applications (P2) and blockchain ecosystems (P4), adhering to the methodology proposed by authors in the publication (Weking et al. 2019).

DAOs were first conceptualized around 2016, and a notably early instance was the "The DAO" project. It was a decentralized venture capital fund managed by a community purely through Ethereum smart contract code. Investment decisions were made based on token holder voting. Although the "The DAO" project ended in disaster due to security vulnerabilities, it drew attention to the potential of decentralized decision-making and governance rooted in the transparency and democratic nature of blockchain technology.

Since then, interest in DAOs and their potential applications has only grown. At their core, DAOs are organizations directly managed by network participants, with decision-making based on collective voting principles. They offer a novel approach to governance, which could stand as an alternative to traditional hierarchical structures, replacing conventional contractual and relational management with blockchain governance (Lumineau et al. 2021).

The application of blockchain technology in DAOs enables the crafting of new organizational structures, where the entire management process, from decision-making to its execution, is encapsulated within a network of automated smart contracts. DAO members interact with these contracts, propose initiatives, and vote on them. A member's voting weight typically hinges on the amount of tokens they possess, ensuring transparency and democratization in the governance process. The distinction between shareholders, managers, and other stakeholders and industry participants becomes blurred, bringing forth numerous advantages (and challenges) (Bellavitis et al. 2022).

Such an organizational structure can be applied not only in the decentralized finance (DeFi) sector but also in a broad spectrum of other industries. According to data from the DeepDAO web resource, there are DAOs operating in areas such as art, culture, gaming, infrastructure, investment, media communication, NFTs, physical asset management, research and data, and venture financing.

However, it's worth noting that not all DAOs use tokens to manage their structure, and not all DAOs have ecosystemic business models. In some cases, DAOs simply represent an organizational form, often used in decentralized finance.

As the popularity of DAOs has grown, some countries have begun to adapt to this new organizational form by offering legal frameworks for their registration. At the time of writing this article, the most favored jurisdictions for DAO registration include the Marshall Islands, the USA (Wyoming), Switzerland, the Cayman Islands, Liechtenstein, Singapore, Panama, the British Virgin Islands, Gibraltar, and the Bahamas.

While smart contract-based governance is a hallmark of DAOs, this form of governance can lead to inefficiencies related to coordination. For instance, the fact that each decision must be voted on by DAO members might take more time than traditional top-down decision-making by managers. Consequently, a voting-based governance structure has limitations when it comes to making time-sensitive decisions (Bellavitis et al. 2022). Research has indicated that the nature of DAO voting is such that of all proposals brought to a vote, 44%, 22%, and 6% were related to grants, new members, and donations, respectively, with the remainder concerning other matters (Bellavitis et al, 2022). This suggests that, as of now, the majority of issues DAO members vote on aren't strategically significant management problems.

Nevertheless, the benefits offered by the decentralization and democratization of governance processes in DAOs have made them attractive to many industries. We can expect that the development of DAOs as increasingly comprehensive governance structures will continue in various sectors, including the financial one.

1.4. DAO VC: A New Perspective on Venture Capital

As previously discussed, the first DAO was created as an innovative form of venture capital fund. This underscores the pivotal role DAO VCs play in the cryptocurrency world and their potential to radically shift traditional investment paradigms.

The primary distinction between DAO VCs and classic venture funds lies in their investment strategy: While traditional VCs typically acquire equity in a startup and await its long-term growth, pre-planning an exit strategy, DAO VCs adopt a different tactic. Instead of taking direct ownership in the startup, they seek a return on investment (ROI) via the appreciation of a token's value as the startup demonstrates growth and success (Anand, Chauhan, 2020; Andreas, Espen, 2023). This method offers token investors flexibility to exchange their investments for fiat currency or other cryptocurrencies on exchanges, in contrast to the often lengthy and challenging exit strategies of traditional VCs (Momtaz, 2021; Zalan, 2018; Andreas, Espen, 2023).

Another substantial difference between DAO VCs and traditional VCs emerges not just in their investment strategy but in the method of fundraising. They employ blockchain technology to gather funds from a decentralized network of investors. This approach enables global investors to invest in early-stage startups, while the startups receive funding via a more streamlined regulatory procedure by selling tokens (or via SAFT contracts). It also reduces the entry barrier for retail investors, granting them an opportunity to participate in the early financing of promising projects. Such ingenuity emphasizes the decentralized and global nature of today's investment landscape.

However, it's essential to acknowledge the risks associated with DAO VCs. Foremost among these challenges are potential vulnerabilities in the smart contracts governing the DAOs, legal and regulatory hurdles, and the imperative for proper evaluation of startups — issues that participants in such decentralized organizations might face.

Recent research confirms the significance of crypto funds in the industry. Scholars have observed that crypto funds positively impact the financial outcomes of token-supported startups (Dombrowski et al. 2023). These findings offer valuable insights into how DAO VCs and crypto funds at large may herald the next generation of innovative investments.

In conclusion, DAO VCs represent a potent tool in today's investment world. They present novel opportunities for startups and investors, signifying a paradigm shift in how investments are gathered, managed, and executed. They also allow both parties to find more flexible and innovative collaboration pathways, primarily within the cryptocurrency market realm.

## Research Methodology

### 2.1. Metrics and Anasis Tools Used

The goal of our study is to determine the relationship between decentralization in DAO governance and the financial efficiency of these organizations in the decentralized finance (DeFi) sector. To do this, we employ tools and metrics utilized in preceding research to determine the level of decentralization and centralization of blockchain infrastructure, specifically the Gini coefficient (Kusmierz, Overko, 2022). This coefficient reflects the degree of token distribution inequality among its holders. A value tending to 1 indicates a high degree of centralization, while a value tending to zero signifies decentralization. Renowned among economists, this coefficient is chosen as the most accurate predictor of decentralization and democratic practices in the case of tokenized DAO governance.

We calculate the Gini coefficient for 32 companies across 9 DeFi sub sectors using a Python script, based on a formula proposed by researchers (Ellipsix, 2013). Notably, our script preemptively removes smart contracts (money pools not owned by any single entity) from the wallet data of each analyzed token's owners. To refine our analysis, we use Python to split (sorted from the largest to smallest holder) token owner wallets into two groups, each equal in wallet address count, and calculate Gini coefficients for each group (LnGC and LnGD). For an in-depth analysis of distribution among the top holders (LnGC), we further segment this group into two subgroups (LnGE and LnGF), ensuring an equal aggregate token quantity in each, and likewise compute their Gini coefficients.

Subsequently, we employ a linear regression analysis method (robust) to discern the relationship between the Gini coefficient (G) and the annual return on investment (ROI vs DeFi 1Y) in DeFi companies. We also invoke the natural logarithm interpretation rule in regression (Stata, 2023) to understand how a 1% change in governance decentralization correlates with a change in ROI. The findings are presented as a recommendation for the academic community and practitioners in the domain of organizations using blockchain and decentralized decision-making mechanisms.

### 2.2. Sample Examined

The research sample comprises 32 representative companies from each of the 9 DeFi sectors previously discussed, as per the Coingecko resource classification. The company selection is grounded on randomness and representativeness of each DeFi sector.

A pivotal criterion for including companies in the sample is their DAO status, i.e., the presence of a token on the blockchain used for voting and decision-making. Information regarding DAO status was confirmed using resources such as Snapshot, Deepdao, and Messari.

Throughout the analysis, data about token holder wallets were gathered and utilized. These details were sourced from blockchain analytics platforms such as Etherscan, Arbiscan, and BscScan. The collated data encompasses information about the number of token holders, the volume of assets owned by each user, and the quantity of tokens in circulation. This data facilitates our examination of the degree of decentralization in token ownership for the selected companies.

## Analysis of the Relationship between Decentralization and Financial Efficiency

### 3.1. Interpretation of Acquired Data

The study was conducted among 32 DeFi companies. Primarily, we divided the data on the wallets of the token holders of each company into two groups, equal in terms of address count, using data sorted from the largest to the smallest holder. Based on these groups, Gini coefficients were computed for the first (C) and second (D) groups.

Subsequently, we segmented the first address group by the volume of tokens into two finer subgroups, so that the first (E) and second (F) had an equal sum of tokens. This was executed for a more detailed analysis of token distribution between large and small holders, within the most significant group (C). Following this, Gini coefficients were calculated for groups E and F respectively. The coefficient calculation outcomes are presented in Table 1 (refer to Appendix 1).

Throughout the regression analysis, we employ the interpretation rule of the natural logarithm (Ln) of key variables, allowing us to interpret the beta coefficients of the regression as elasticity coefficients.

For analyzing the selected variables LnGC, LnGD, LnGE, and LnGF in relation to the ROI of companies, we utilized robust linear regression. This method demonstrates resilience to outlier values, which is essential when dealing with data where there might be substantial fluctuations, such as data associated with cryptocurrencies.

The results of the regression analysis are provided in the table below.

**Table 4.** Analysis Results of LnROI in relation to LnGC, LnGD, LnGE, and LnGF

|  | OLS (1) | OLS (2) | OLS (3) | OLS (4) |
|---|---|---|---|---|
| LnGC | -5.872964 | - | - | - |
|  | (6.658348) |  |  |  |
| LnGD | -0.1102615** | -0.1082782** | -0.1107804** | -0.117077** |
|  | (0.0377272) | (0.0367026) | (0.0382616) | (0.0371044) |
| LnGE | 0.0418288 | 0.0407845 | 0.0401596 | - |
|  | (0.035893) | (0.0352715) | (0.0342273) |  |
| LnGF | 3.060148 | 0.057727 | - | - |
|  | (3.299138) | (0.5912853) |  |  |
| Constant | -0.3577703* | -0.3478227* | -0.3579189* | -0.4456323*** |
|  | (0.1479532) | (0.1440672) | (0.1485699) | (0.1270005) |
| Observations | 32 | 32 | 32 | 32 |

*Notes: We employ OLS regression with robust standard errors. The number in parentheses represents the standard error. Asterisks indicate the significance level of the t-test (* $p \leq 0.05$, ** $p \leq 0.01$, *** $p \leq 0.001$).*

According to our findings, the variable LnGD demonstrated a statistically significant negative relationship with LnROI across all four regression models. This suggests that the level of inequality in token distribution among the smaller holders (second address group) impacts the ROI of DeFi companies. It's worth noting that the significance level (p-value) falls within the 0.004-0.007 range, affirming the statistical relevance of this outcome. The other variables (LnGC, LnGE, and LnGF) did not indicate a statistically significant correlation with LnROI.

The beta coefficient for LnGD in the final model is -0.117077. Since we employed the natural logarithm (Ln) for the regression, this coefficient can be construed as an elasticity coefficient. This denotes that a 1% decrease in LnGD (i.e., a 1% rise in the uniformity of token distribution among holders in the second group) is associated with an approximate 0.117% augmentation in LnROI.

### 3.2. Findings and Suggestions for Further Research

During the course of this study, it was determined that a more even distribution of tokens among holders from the second group leads to higher investment returns. This observation is particularly relevant in the context of the elasticity effect, identified within the usage of the natural logarithm of variables in the conducted regression analyses. Therefore, it was discovered that a 1% increase in the Gini coefficient of the second group (LnGD) is associated with a decrease in the magnitude of LnROI by 0.11 - 0.12%.

These findings can be invaluable for researchers and practitioners in the field of crypto-economics and DeFi, as they point to the significance of a more balanced distribution of tokens among holders to enhance investment returns. In this context, strategies and mechanisms promoting such distribution might be useful for improving the profitability and stability of DeFi projects.

However, it's essential to note that our model can be supplemented and refined. Our results did not indicate a statistically significant influence of variables LnGE and LnGF (Gini coefficients for major holders) on LnROI. This might suggest that the concentration of tokens among large holders doesn't substantially impact investment profitability. But this should not diminish the importance of further research in this direction, as the balance between large and small holders remains a crucial aspect of the DeFi ecosystem.

In light of the obtained results, it's proposed to conduct further studies aiming to understand how other factors influence the financial efficiency of DeFi companies. Perhaps considering parameters such as user count, Total Value Locked, trading volume, DAO offering quality, and overall company management quality might lead to a deeper understanding of how high efficiency metrics are achieved in this domain.

## Conclusion

This research was conducted with the objective of identifying the relationship between token distribution in DeFi projects and their profitability. Specifically, we focused on determining how various levels of token concentration impact investment returns (ROI).

Using data from 32 DeFi companies, we performed multiple linear regressions and found that the degree of token concentration among smaller holders (LnGD) has a significant effect on investment returns (LnROI). According to our findings, a uniform distribution of tokens among smaller holders leads to higher investment returns. However, we did not find a statistically significant influence of token distribution among larger holders on investment profitability.

These results hold significant implications for practitioners and researchers in the DeFi sector, as they highlight the importance of evenly distributing tokens among smaller holders to achieve higher investment returns. Nonetheless, these findings also pave the way for further research. It would be beneficial to determine how other factors, such as overall management quality and token distribution strategies, impact investment profitability.

In conclusion, this study serves as a crucial step in understanding the dynamics of token distribution in DeFi projects and their influence on the financial efficiency of companies. We anticipate that our conclusions and suggestions will assist researchers and practitioners in this promising domain in the future.

## Bibliography

1. Al Ajlouni, A.T., Al-Hakim, M.S. (2018). Financial Technology in Banking Industry: Challenges and Opportunities. DOI: 10.2139/ssrn.3340363
2. Davidson, S., De Filippi, P., Potts, J. (2017). Blockchains and the Economic Institutions of Capitalism. *Journal of Institutional Economics 14(4):1-20.* DOI: 0.1017/S1744137417000200
3. Coingecko. (2022). Annual Crypto Industry Report. https://assets.coingecko.com/reports/2022/CoinGecko-2022-Annual-Crypto-Industry-Report.pdf
4. IMF. (2016). Monetary and financial statistics manual and compilation guide. pp. 32-33. https://www.imf.org › mfsmcg_merged-web-pdf
5. Aleshina, A. (2022). The impact of financial technologies and decentralized finance (DeFi) on threats to the infrastructure of the national economy. *Financial markets and banks. 1:121-125.* https://cyberleninka.ru/article/n/vozdeystvie-finansovyh-tehnologiy-i-detsentralizovannyh-finansov-defi-na-ugrozy-infrastrukture-natsionalnoy-ekonomiki
6. Weking, J., Mandalenakis, M., Hein, A., Hermes, S., Böhm, M., Krcmar, H. (2019). The impact of blockchain technology on business models – a taxonomy and archetypal patterns.

https://www.researchgate.net/publication/338118210_The_impact_of_blockchain_technology_on_business_models_-_a_taxonomy_and_archetypal_patterns

7. Lumineau, F., Wang, W., Schilke, O. (2021). Blockchain Governance – A New Way of Organizing Collaborations?. *Organization Science 32(2):500-521.* https://doi.org/10.1287/orsc.2020.1379

8. Bellevitis, C., Fisch, C., Momtaz P.P. (2022). The rise of decentralized autonomous organizations (DAOs): a first empirical glimpse. https://www.researchgate.net/publication/359712137_The_rise_of_decentralized_autonomous_organizations_DAOs_a_first_empirical_glimpse

9. Andreas, S., Espen, B. (2023). Best Practices in Decentralized Autonomous Organization (DAO) Venture Capital. https://uia.brage.unit.no/uia-xmlui/bitstream/handle/11250/3081668/no.uia%3Ainspera%3A143804064%3A35492471.pdf?sequence=1

10. Dombrowski, N., Wolfgang, D., Momtaz., P.P. (2023). Performance measurement of crypto funds. https://www.sciencedirect.com/science/article/pii/S016517652300143X#fig2

11. Kusmierz, B., Overko R. (2022). How centralized is decentralized? Comparison of wealth distribution in coins and tokens. https://arxiv.org/pdf/2207.01340.pdf

*Online sources:*

12. Gini Coefficient formula. URL: https://www.ellipsix.net/blog/2012/11/the-gini-coefficient-for-distribution-inequality.html

13. Interpretation of logarithms in a regression. URL: https://www.princeton.edu/~otorres/Stata/inference.htm

14. Analytical web resource Coingecko. URL: https://www.coingecko.com/

15. Analytical web resource Messari. URL: https://messari.io/

16. Analytical web resource Etherscan. URL: https://etherscan.io/

Appendix

**Appendix 1.**

**Table 1.** Gini Coefficients of DeFi Companies

| Company | Governance token | Gini for first group (C) | Gini for second group (D) | Gini for top 50% of first group (E) | Gini for bottom 50% of first group (F) |
|---|---|---|---|---|---|
| Uniswap | UNI | 0,98 | 0,19 | 0,27 | 0,96 |
| Curve | CRV | 0,98 | 0,23 | 0,41 | 0,96 |
| Trader Joe | JOE | 0,99 | 0,26 | 0,00 | 0,99 |
| Convex | CVX | 0,94 | 0,25 | 0,40 | 0,89 |
| Alpha | ALPHA | 0,98 | 0,24 | 0,17 | 0,97 |
| Yearn finance | YFI | 0,98 | 0,25 | 0,27 | 0,96 |
| Beefy finance | BIFI | 0,99 | 0,18 | 0,01 | 0,99 |
| Lido DAO | LDO | 0,99 | 0,26 | 0,43 | 0,97 |
| Rocket pool | RPL | 0,92 | 0,24 | 0,41 | 0,85 |
| Frax finance | FXS | 0,96 | 0,25 | 0,44 | 0,93 |
| Ankr | ANKR | 0,99 | 0,19 | 0,29 | 0,98 |
| Chainlink | LINK | 0,96 | 0,24 | 0,43 | 0,91 |
| Band protocol | BAND | 0,98 | 0,08 | 0,08 | 0,97 |
| Uma | UMA | 0,99 | 0,26 | 0,32 | 0,98 |
| Api3 | API3 | 0,94 | 0,24 | 0,37 | 0,88 |
| AAve | AAVE | 0,95 | 0,26 | 0,37 | 0,90 |
| Maker | MKR | 0,99 | 0,23 | 0,37 | 0,98 |
| Compound | COMP | 0,99 | 0,24 | 0,25 | 0,99 |
| Synthetix | SNX | 0,97 | 0,24 | 0,31 | 0,94 |
| GMX | GMX | 1,00 | 0,01 | 0,51 | 0,99 |
| dYdX | DYDX | 0,99 | 0,26 | 0,46 | 0,98 |
| Perpetual | PERP | 0,98 | 0,22 | 0,00 | 0,97 |
| Insure DeFi | SURE | 0,99 | 0,24 | 0,07 | 0,99 |
| Etherisc Dip | DIP | 0,90 | 0,24 | 0,48 | 0,81 |
| SwissBorg | CHSB | 0,85 | 0,26 | 0,66 | 0,70 |
| Dexe | DEXE | 0,99 | 0,15 | 0,06 | 0,98 |
| Enzyme | MLN | 0,98 | 0,26 | 0,33 | 0,96 |
| DefiPulse | DPI | 0,84 | 0,27 | 0,58 | 0,71 |
| Hifi finance | HIFI | 0,87 | 0,24 | 0,21 | 0,79 |
| BarnBridge | BOND | 0,98 | 0,23 | 0,34 | 0,96 |
| 88mph | MPH | 0,92 | 0,23 | 0,44 | 0,85 |
| Notional | NOTE | 0,88 | 0,25 | 0,36 | 0,80 |